\documentclass{ws-procs975x65}%
\usepackage{amsmath}
\usepackage{amsfonts}
\usepackage{amssymb}
\usepackage{graphicx}%
\def\beq{\begin{equation}}
\def\eeq{\end{equation}}
\begin{document}

\title{The Distorted Wheeler-DeWitt Equation}
\author{Remo Garattini}

\begin{abstract}
The Wheeler-DeWitt Equation represents a tool to study Quantum Gravity and
Quantum Cosmology. Its solution in a very general context is, of course,
impossible. To this purpose we consider some distortions of General Relativity
like Gravity's Rainbow, Varying Speed of Light Cosmology, Generalized
Uncertainty Principle deformations and Ho\v{r}ava-Lifshitz gravity which could
allow the calculation of some observables like the cosmological constant. For
simplicity we consider only the Mini-Superspace approach related to a
Friedmann-Lema\^{\i}tre-Robertson-Walker space-time.

\end{abstract}

\bigskip

\address{Universit\`{a} degli Studi di Bergamo, \\
Dipartimento di Ingegneria e scienze applicate,\\
Viale Marconi,5 24044 Dalmine (Bergamo) ITALY\\
I.N.F.N. - sezione di Milano, Milan, Italy\\
$^*$E-mail:remo.garattini@unibg.it}

\bodymatter

\section{Introduction}

\label{p1}

The Wheeler-DeWitt (WDW) equation represents the quantum version of the
classical constraint describing the invariance with respect to time
reparametrization\cite{DeWitt}. In the context of the
Friedmann-Lema\^{\i}tre-Robertson-Walker space-time, it assumes the simple
form
\begin{align}
H\Psi\left(  a\right)   &  =\left[  -a^{-q}\left(  \frac{\partial}{\partial
a}a^{q}\frac{\partial}{\partial a}\right)  +\frac{9\pi^{2}}{4G^{2}}\left(
a^{2}-\frac{\Lambda}{3}a^{4}\right)  \right]  \Psi\left(  a\right)
\nonumber\\
&  =\left[  -\frac{\partial^{2}}{\partial a^{2}}-\frac{q}{a}\frac{\partial
}{\partial a}+\frac{9\pi^{2}}{4G^{2}}\left(  a^{2}-\frac{\Lambda}{3}%
a^{4}\right)  \right]  \Psi\left(  a\right)  =0,\label{WDW_0}%
\end{align}
where we have introduced the following line element%
\begin{equation}
ds^{2}=-N^{2}dt^{2}+a^{2}\left(  t\right)  d\Omega_{3}^{2}\label{FRW}%
\end{equation}
and where we have denoted with $d\Omega_{3}^{2}=\gamma_{ij}dx^{i}dx^{j}$ the
metric on the three-sphere. $N$ is the lapse function, $a(t)$ is the scale
factor, $G$ and $\Lambda$ are the Newton's constant and the cosmological
constant, respectively. We have also introduced a factor order ambiguity $q$.
In Eq.$\left(  \ref{WDW_0}\right)  $ all the degrees of freedom except the
scale factor $a\left(  t\right)  $ have been integrated and matter fields have
been excluded. If the WDW equation is interpreted as an eigenvalue equation,
one simply finds%
\begin{equation}
H\Psi\left(  a\right)  =E\Psi\left(  a\right)  =0,
\end{equation}
namely a zero energy eigenvalue. However, it appears that the WDW equation has
also a hidden structure. Indeed Eq.$\left(  \ref{WDW_0}\right)  $ has the
structure of a Sturm-Liouville eigenvalue problem with the cosmological
constant interpreted as the associated eigenvalue. We recall to the reader
that a Sturm-Liouville differential equation is defined by%
\begin{equation}
\frac{d}{dx}\left(  p\left(  x\right)  \frac{dy\left(  x\right)  }{dx}\right)
+q\left(  x\right)  y\left(  x\right)  +\lambda w\left(  x\right)  y\left(
x\right)  =0\label{SL}%
\end{equation}
and the normalization is defined by%
\begin{equation}
\int_{a}^{b}dxw\left(  x\right)  y^{\ast}\left(  x\right)  y\left(  x\right)
,
\end{equation}
where the boundary conditions are momentarily suspended. It is a standard
procedure, to convert the Sturm-Liouville problem $\left(  \ref{SL}\right)  $
into a variational problem of the form%
\begin{equation}
F\left[  y\left(  x\right)  \right]  =\frac{-\int_{a}^{b}dxy^{\ast}\left(
x\right)  \left[  \frac{d}{dx}\left(  p\left(  x\right)  \frac{d}{dx}\right)
+q\left(  x\right)  \right]  y\left(  x\right)  }{\int_{a}^{b}dxw\left(
x\right)  y^{\ast}\left(  x\right)  y\left(  x\right)  }\,,\label{Funct}%
\end{equation}
with unspecified boundary condition. If $y\left(  x\right)  $ is an
eigenfunction of $\left(  \ref{SL}\right)  $, then%
\begin{equation}
\lambda=\frac{-\int_{a}^{b}dxy^{\ast}\left(  x\right)  \left[  \frac{d}%
{dx}\left(  p\left(  x\right)  \frac{d}{dx}\right)  +q\left(  x\right)
\right]  y\left(  x\right)  }{\int_{a}^{b}dxw\left(  x\right)  y^{\ast}\left(
x\right)  y\left(  x\right)  }\,,
\end{equation}
is the eigenvalue, otherwise%
\begin{equation}
\lambda_{1}=\min_{y\left(  x\right)  }\frac{-\int_{a}^{b}dxy^{\ast}\left(
x\right)  \left[  \frac{d}{dx}\left(  p\left(  x\right)  \frac{d}{dx}\right)
+q\left(  x\right)  \right]  y\left(  x\right)  }{\int_{a}^{b}dxw\left(
x\right)  y^{\ast}\left(  x\right)  y\left(  x\right)  }\,.
\end{equation}
\textbf{\ }The minimum of the functional $F\left[  y\left(  x\right)  \right]
$ corresponds to a solution of the Sturm-Liouville problem $\left(
\ref{SL}\right)  $ with the eigenvalue $\lambda.$ In the case of the FLRW
model we have the following correspondence%
\begin{align}
p\left(  x\right)   &  \rightarrow a^{q}\left(  t\right)  \,,\nonumber\\
q\left(  x\right)   &  \rightarrow\left(  \frac{3\pi}{2G}\right)  ^{2}%
a^{q+2}\left(  t\right)  \,,\nonumber\\
w\left(  x\right)   &  \rightarrow a^{q+4}\left(  t\right)  \,,\nonumber\\
y\left(  x\right)   &  \rightarrow\Psi\left(  a\right)  \,,\nonumber\\
\lambda &  \rightarrow\frac{\Lambda}{3}\left(  \frac{3\pi}{2G}\right)  ^{2}\,.
\end{align}
Since $a\left(  t\right)  \in\left[  0,\infty\right)  $, the normalization
becomes%
\begin{equation}
\int_{0}^{\infty}daa^{q+4}\Psi^{\ast}\left(  a\right)  \Psi\left(  a\right)
,\label{Norm1}%
\end{equation}
where it is understood that $\Psi\left(  \infty\right)  =0$. In the
Mini-Superspace approach with a FLRW background, one finds%
\begin{equation}
\frac{\int\mathcal{D}aa^{q}\Psi^{\ast}\left(  a\right)  \left[  -\frac
{\partial^{2}}{\partial a^{2}}-\frac{q}{a}\frac{\partial}{\partial a}%
+\frac{9\pi^{2}}{4G^{2}}a^{2}\right]  \Psi\left(  a\right)  }{\int
\mathcal{D}aa^{q+4}\Psi^{\ast}\left(  a\right)  \Psi\left(  a\right)  }%
=\frac{3\Lambda\pi^{2}}{4G^{2}}.\label{WDW_1}%
\end{equation}
As a concrete case, fixing $q=0$ and taking as a trial wave function
$\Psi\left(  a\right)  =\exp\left(  -\beta a^{2}\right)  $, one finds
$\Psi\left(  a\right)  \rightarrow0$ when $a\rightarrow\infty$. Then the only
solution allowed is complex and therefore it must be discarded\cite{RemoHL}.
Of course, the general $q$ case is much more complicated\cite{RGMdL}. Note
that the global energy eigenvalue is still vanishing. What we can compute in
the Sturm-Liouville formulation is the degree of degeneracy which is
represented by the cosmological constant and the value of the cosmological
constant itself. In the next section we give the general guidelines to build a
Sturm-Liouville problem associated to the WDW equation when General Relativity
(GR) is distorted. Units in which $\hbar=c=k=1$ are used throughout the paper.

\section{The Mini-Superspace Approach to the Cosmological Constant in
Distorted Quantum Cosmology}

In this section we will give some glimpses on how the WDW is modified when we
distort GR. The reference metric will be the FLRW line element and we will
consider modifications coming from Gravity's Rainbow, HL\ Gravity, VSL
cosmology and GUP deformation. We begin to consider Gravity's Rainbow and HL Gravity.

\subsection{Gravity's Rainbow and HL Gravity}

Gravity's Rainbow is a distortion of the metric tensor around and beyond the
Planck scale\cite{MagSmo}. The basic ingredient is the definition of two
arbitrary functions $g_{1}\left(  E/E_{\mathrm{Pl}}\right)  $ and
$g_{2}\left(  E/E_{\mathrm{Pl}}\right)  $, which have the following property%
\begin{equation}
\lim_{E/E_{\mathrm{Pl}}\rightarrow0}g_{1}\left(  E/E_{\mathrm{Pl}}\right)
=1\qquad\mathrm{and}\qquad\lim_{E/E_{\mathrm{Pl}}\rightarrow0}g_{2}\left(
E/E_{\mathrm{Pl}}\right)  =1.\label{lim}%
\end{equation}
For a FLRW metric, the deformation acts in the following way,%
\begin{equation}
ds^{2}=-\frac{N^{2}\left(  t\right)  }{g_{1}^{2}\left(  E/E_{\mathrm{Pl}%
}\right)  }dt^{2}+\frac{a^{2}\left(  t\right)  }{g_{2}^{2}\left(
E/E_{\mathrm{Pl}}\right)  }d\Omega_{3}^{2}~,\label{FRWMod}%
\end{equation}
where%
\begin{equation}
d\Omega_{3}^{2}=\gamma_{ij}dx^{i}dx^{j}\label{domega}%
\end{equation}
is the line element on the three-sphere. Of course, ordinary gravity is
recovered when $E/E_{\mathrm{Pl}}\rightarrow0$. When we generalize Eq.$\left(
\ref{WDW_1}\right)  $ to include $g_{1}\left(  E/E_{\mathrm{Pl}}\right)  $ and
$g_{2}\left(  E/E_{\mathrm{Pl}}\right)  $, the equation modifies in the
following way%
\begin{equation}
\left[  -\frac{\partial^{2}}{\partial a^{2}}-\frac{q}{a}\frac{\partial
}{\partial a}+U\left(  a,E/E_{\mathrm{Pl}}\right)  \right]  \Psi\left(
a\right)  =0,\label{WDWGRw}%
\end{equation}
where we have set $N=1$ and defined the distorted potential as%
\begin{equation}
U\left(  a,E/E_{\mathrm{Pl}}\right)  =\left[  \frac{3\pi g_{2}\left(
E/E_{\mathrm{Pl}}\right)  }{2Gg_{1}\left(  E/E_{\mathrm{Pl}}\right)  }\right]
^{2}a^{2}\left[  1-\frac{a^{2}}{a_{0}^{2}g_{2}^{2}\left(  E/E_{\mathrm{Pl}%
}\right)  }\right]  ~,\label{U(a,E)}%
\end{equation}
with $a_{0}=\sqrt{3/\Lambda}$. The potential $U\left(  a,E/E_{\mathrm{Pl}%
}\right)  $ has been used to discuss a possible alternative explanation of the
inflation problem\cite{RGMS}. Moreover, always in the context of the FLRW
space-time, it is possible to build a bridge between Gravity's Rainbow and HL
theory\cite{RGENS}. Indeed, if one considers%
\begin{equation}
g_{1}\left(  E/E_{P}\right)  \equiv g_{1}\left(  E\left(  a\left(  t\right)
\right)  /E_{P}\right)  \qquad g_{2}\left(  E/E_{P}\right)  \equiv
g_{2}\left(  E\left(  a\left(  t\right)  \right)  /E_{P}\right)
\end{equation}
with the choice%
\begin{equation}
g_{1}^{2}\left(  E\left(  a\left(  t\right)  \right)  /E_{P}\right)
f\!\left(  A\left(  t\right)  ,a\right)  =1
\end{equation}
and%
\begin{equation}
g_{2}^{2}\left(  E\left(  a\left(  t\right)  \right)  /E_{P}\right)  \frac
{6}{a^{2}\left(  t\right)  }=\frac{6}{a^{2}\left(  t\right)  }\left[
1-\frac{2\kappa b}{a^{2}\left(  t\right)  }-\frac{4\kappa^{2}c}{a^{4}\left(
t\right)  }\right]  ,
\end{equation}
one finds that the WDW equation becomes
\begin{equation}
-\frac{\partial^{2}}{\partial a^{2}}+\frac{\left(  3\lambda-1\right)  }%
{\kappa^{2}}24\pi^{4}a^{4}\left(  t\right)  \left[  \frac{6}{a^{2}\left(
t\right)  }-\frac{12\kappa b}{a^{4}\left(  t\right)  }-\frac{24\kappa^{2}%
c}{a^{6}\left(  t\right)  }-2\Lambda\right]  \Psi\left(  a\right)
=0.\label{HHL}%
\end{equation}
To obtain Eq.$\left(  \ref{HHL}\right)  $, we have also defined%
\begin{equation}
A\left(  t\right)  =\frac{1}{g_{2}\left(  E\left(  a\left(  t\right)  \right)
/E_{P}\right)  E_{P}}\frac{d}{dE}\left[  g_{2}\left(  E\left(  a\left(
t\right)  \right)  /E_{P}\right)  \right]  \frac{dE}{da}%
\end{equation}
and%
\begin{equation}
f\!\left(  A\left(  t\right)  ,a\right)  =\left[  1-2a\left(  t\right)
A\left(  t\right)  +A^{2}\left(  t\right)  a\left(  t\right)  ^{2}\right]  .
\end{equation}
Moreover, by identifying%
\begin{align}
&  3g_{2}+g_{3}=b\nonumber\\
&  9g_{4}+3g_{5}+g_{6}=c,\label{bcgs}%
\end{align}
it is immediate to recognize that Eq.$\left(  \ref{HHL}\right)  $ represents
the WDW equation for the projectable version of a Ho\v{r}ava-Lifshitz gravity,
without detailed balanced condition.

\subsection{VSL Cosmology}

A VSL cosmology model is described by the following line element%
\begin{equation}
ds^{2}=-N^{2}\left(  t\right)  c^{2}\left(  t\right)  dt^{2}+a^{2}\left(
t\right)  d\Omega_{3}^{2}, \label{FRWc}%
\end{equation}
and where $c\left(  t\right)  $ is an arbitrary function of time with the
dimensions of a $\left[  length/time\right]  $. Following
\cite{Barrow1,Barrow2,Barrow3}, we assume that%
\begin{equation}
c\left(  t\right)  =c_{0}\left(  \frac{a\left(  t\right)  }{a_{0}}\right)
^{\alpha} \label{c(t)}%
\end{equation}
where $a_{0}$ is a reference length scale. The form of the background is such
that the shift function $N^{i}$ vanishes. In this case, when $\pi_{a}$ is
promoted to an operator, we can write%
\begin{equation}
\pi_{a}^{2}\rightarrow-\left(  \hbar c\left(  t\right)  \right)  ^{2}%
a^{-q}\frac{\partial}{\partial a}a^{q}\frac{\partial}{\partial a},
\end{equation}
where we have introduced a factor ordering ambiguity. Thus the WDW equation
$\mathcal{H}\Psi=0$ simply becomes%
\begin{equation}
\left(  -\frac{\partial^{2}}{\partial a^{2}}-\frac{q}{a}\frac{\partial
}{\partial a}+U_{c}\left(  a\right)  \right)  \Psi\left(  a\right)  =0,
\label{WDWg}%
\end{equation}
where we have assumed that the factor ordering is not distorted by the
presence of a VSL and we have set $N=1$. The quantum potential is defined as%
\begin{equation}
U_{c}\left(  a\right)  =\left(  \frac{3\pi}{2G\hbar}\right)  ^{2}a^{2}%
c^{6}\left(  t\right)  \left(  1-\frac{\Lambda}{3}a^{2}\right)  =\left(
\frac{3\pi c_{0}^{3}}{2G\hbar a_{0}^{3\alpha}}\right)  ^{2}a^{2+6\alpha
}\left(  1-\frac{\Lambda}{3}a^{2}\right)  . \label{Uac}%
\end{equation}
Note that the potential $U_{c}\left(  a\right)  $ vanishes in the same points
where the original $U\left(  a\right)  $ has its roots.

\subsection{GUP Deformation}

The WDW equation deformed by a GUP can be derived modifying the original
undeformed momentum conjugate to the scale factor $a\left(  t\right)  $%
\begin{equation}
\tilde{\pi}_{a}=-i\frac{d}{da},
\end{equation}
into\cite{MF}
\begin{equation}
\pi_{a}=\tilde{\pi}_{a}(1-\alpha||\tilde{\pi}_{a}||+2\alpha^{2}||\tilde{\pi
}_{a}||^{2}),
\end{equation}
where $\alpha$ is a coefficient with inverse momentum dimensions, in order to
reestablish the correct powers. Then the WDW equation for a FLRW metric
deformed by a GUP becomes%
\begin{equation}
\left[  \tilde{\pi}_{a}^{2}-2\alpha\pi_{a}^{3}+5\alpha^{2}\pi_{a}^{4}+\left(
\frac{3\pi}{2l_{P}^{2}}\right)  ^{2}a^{2}\left(  1-\frac{\Lambda}{3}%
a^{2}\right)  \right]  \Psi\left(  a\right)  =0.
\end{equation}
This deformation of the WDW equation prevents the existence of singularities
\cite{MF}. This is because this deformation modifies the uncertainty principle
as%
\begin{equation}
\Delta a\Delta\pi_{a}=1-2\alpha<\pi_{a}>+4\alpha^{2}<\pi_{a}^{2}>.
\end{equation}
Thus, we obtain a minimum value for the scale factor of the universe, $\Delta
a\geq\Delta a_{min}$. Note that the action of the GUP is only on the kinetic term.

\section{Summary}

In this work, we have considered the WDW equation on a FLRW background and we
have shown how such an equation is modified when some deviations from GR are
considered. Even if we have only considered a Mini-Superspace approach with
the scale factor $a\left(  t\right)  $ as unique degree of freedom, one has
not to think that this procedure works only on a FLRW background. For
instance, in a series of papers\cite{GaMa,GRw}, it has been shown that
Gravity's Rainbow can keep under control UV divergences, at least to one loop.
This procedure has been widely tested on a spherically symmetric background.
The same procedure can be applied also to Noncommutative geometries\cite{RGPN}
and it has been extended to include also a $f\left(  R\right)  $
theory\cite{Rf(R)}.


\begin{thebibliography}{99}                                                                                               %


\bibitem {DeWitt}B. S. DeWitt, \textsl{Phys. Rev.} \textbf{160}, 1113 (1967).

\bibitem {RemoHL}R. Garattini, \textsl{Phys. Rev. }\textbf{D 86}, 123507
(2012); ArXiv:0912.0136 [gr-qc].

\bibitem {RGMdL}R. Garattini and M. De Laurentis, \textit{The Cosmological
Constant as an Eigenvalue of the Hamiltonian constraint in a Varying Speed of
Light theory}, arXiv:1503.03677 [gr-qc].

\bibitem {MagSmo}J.~Magueijo and L.~Smolin, \textsl{Class.\ Quant.\ Grav.}%
\ \textbf{21}, 1725 (2004). ArXiv: 0305055 [gr-qc].

\bibitem {RGMS}R. Garattini and M. Sakellariadou, \textsl{Phys. Rev.}
\textbf{D} \textbf{90} 043521 (2014), ArXiv:1212.4987 [gr-qc].

\bibitem {RGENS}R. Garattini and E. N. Saridakis, \textsl{Eur.Phys.J. C
}\textbf{75} (2015) 7, 343 ArXiv:1411.7257.

\bibitem {Barrow1}J. D. Barrow,
\textit{Phys. Rev. D} \textbf{59} (1999) 043515 .

\bibitem {Barrow2}J. D. Barrow and J. Magueijo,
\textit{Phys. Lett. B} \textbf{443} (1998) 104 .

\bibitem {Barrow3}J. D. Barrow and J. Magueijo,
\textit{Phys. Lett. B} \textbf{447} (1999) 246.

\bibitem {MF}M.~Faizal, \textsl{Int.\ J.\ Mod.\ Phys.}\ \textbf{A}
\textbf{29}, no. 20, 1450106 (2014) [arXiv:1406.0273 [gr-qc]].

\bibitem {RGMF}R. Garattini and M.~Faizal, \textit{Cosmological Constant from
a Deformation of the Wheeler-DeWitt Equation}, arXiv:1510.04423 [gr-qc].

\bibitem {GaMa}R.~Garattini and G.~Mandanici, \textsl{Phys. Rev.} \textbf{D
83}, 084021 (2011); ArXiv:1102.3803 [gr-qc].

\bibitem {GRw}R.~Garattini, \textsl{Phys.\ Lett.}\ \textbf{B} \textbf{685},
329 (2010), ArXiv: 0902.3927 [gr-qc]. R.~Garattini and G.~Mandanici,
\textsl{Phys. Rev.} \textbf{D 85}, 023507 (2012); ArXiv:1109.6563 [gr-qc].
R.~Garattini, JCAP \textbf{1306}, 017 (2013), ArXiv: 1210.7760 [gr-qc].
R.~Garattini and F.~S.~N.~Lobo, \textsl{Phys.\ Rev.}\ \textbf{D} \textbf{85},
024043 (2012), ArXiv:1111.5729 [gr-qc]. R.~Garattini,
\textsl{Int.\ J.\ Mod.\ Phys.\ Conf.\ Ser.}\ \textbf{14}, 326 (2012),
ArXiv:1112.1630 [gr-qc]. R. Garattini and F.S.N. Lobo, \textsl{Eur. Phys. J.}
\textbf{C 74} (2014), ArXiv:1303.5566 [gr-qc]. R. Garattini and B.
Majumder,\textsl{\ Nucl. Phys.} \textbf{B} \textbf{883} (2014),
ArXiv:1305.3390 [gr-qc].R. Garattini and B. Majumder, \textsl{Nucl. Phys.}
\textbf{B} \textbf{884, }(2014), ArXiv:1311.1747 [gr-qc].

\bibitem {RGPN}R.~Garattini and P.~Nicolini, \textsl{Phys.\ Rev.}\ \textbf{D}
\textbf{83}, 064021 (2011), ArXiv:1006.5418[gr-qc].

\bibitem {Rf(R)}R.~Garattini, \textsl{JCAP} \textbf{1306}, 017 (2013),
ArXiv:1210.7760; S. Capozziello and R. Garattini, \textsl{Class.Quant.Grav.
}\textbf{24}:1627 (2007), ArXiv: gr-qc/0702075.
\end{thebibliography}
\end{document}